\documentstyle[12pt,epsfig,a4]{article}

\begin{document}
\begin{center}
\large {\bf Secure communication with a publicly known key} \\
\vspace*{0.5cm}
Almut Beige,$^1$ Berthold-Georg Englert,$^{1,2}$
Christian Kurtsiefer,$^3$ and Harald Weinfurter$^{1,3}$ \\[.5cm]

\normalsize
$^1$Max-Planck-Institut f\"ur Quantenoptik, Hans-Kopfermann-Str. 1, \\  
85748 Garching, Germany \\
$^2$Departments of Mathematics and Physics, Texas A\&M University, \\
College Station, Texas 77843, U. S. A.\\
$^3$Sektion Physik, Universit\"at M\"unchen, Schellingstrasse 4, 80799 
M\"unchen, Germany \\[0.5cm]
\end{center}

\vspace*{0.5cm}
\begin{abstract}
{We present a scheme for direct and confidential communication between
Alice and Bob, where there is no need for establishing a shared secret
key first, and where the key used by Alice even will become known publicly. 
The communication is based on the exchange of single photons and 
each and every photon transmits one bit of Alice's message without 
revealing any information to a potential eavesdropper.}
\end{abstract}

\vspace*{0.5cm}
\vspace*{1cm}

\noindent {\bf 1. Introduction}
\vspace*{0.5cm}

It is generally believed that cryptography schemes are only completely secure when 
the two communicating parties, {\em Alice} and {\em Bob}, establish a shared secret 
key before the transmission of a message. This means they first have to create a 
random bit sequence, which is not known to anyone else, and which is 
of the same length as the message. In order to communicate, Alice then 
multiplies the bits of the message one by one with the key bits. When she announces 
the result to Bob, or even publicly, then he is the only one who can interpret it 
and deduce Alice's message.

As shown in a seminal paper by Bennett and Brassard in 1984 \cite{BB84}, Alice 
and Bob can establish a shared secret key by exchanging single qubits, physically 
realised by the polarisation of photons, for example.
The protocol of the proposed scheme, that became known as BB84, is as follows. 
First, Alice prepares a photon 
in a certain polarisation state, a basis vector in a two-dimensional Hilbert space. 
Thereby she chooses at random between two complementary bases. Afterwards she forwards 
the photon to Bob who now performs a measurement on the incoming state. If he chooses 
the same basis as Alice, which happens with a chance of $50 \, \%$, they can agree 
about one key bit. At the end of a transmission they check whether it was secure 
or not by just comparing some key bits. An eavesdropper can be noticed because 
his interception causes an error rate of at least $25 \,\%$. 

In 1987 Vaidman, Aharonov and Albert published a paper \cite{VAA} with the 
title ``How to ascertain the values of $\sigma_x$, $\sigma_y$ and $\sigma_z$ of a 
spin-${1 \over 2}$ particle'' and described a paradox that later became known as the
{\em Mean King's Problem} \cite{VAA2}. The quantum-optical version of the king's 
problem proposed recently \cite{Englert} suggests 
a new cryptography scheme, presented in Refs.~\cite{Letter,book}. 
In the present paper we focus on a further development, namely a modification that 
allows Alice to send a message to Bob without the need to establish a shared secret 
key first. 

The protocol of this new scheme has, of course, many similarities to 
BB84 \cite{BB84}, its later modifications \cite{BB92,Bruss,Bour} and the
proposal made by Ekert in 1991 based on entangled photon pairs \cite{Ekert}. 
But it is much more than just another modification. In contrast 
to BB84 and its various ``analytical continuations'', which are {\em probabilistic},
the scheme we describe here is {\em deterministic}. 
Each and every photon sent and detected will eventually contribute a key bit. 
In addition, and this is another important requirement for direct and confidential 
communication, no information is revealed to a potential eavesdropper. 
The only other proposal with deterministic features is the one of Goldenberg 
and Vaidman \cite{Goldenberg}.

Altogether, the transmission of a message 
becomes more efficient than in other schemes.
The price to be paid for this efficiency raise is that each photon now has to be
prepared in a two-qubit state and not only in a single-qubit state. To obtain these 
states Alice can use, for instance, the spatial binary alternative of a photon
with the basis states $|{\sf R} \rangle$ and $|{\sf L} \rangle$ and the two 
polarisations $|{\sf v} \rangle$ and $|{\sf h} \rangle$. Here, $|{\sf R} \rangle$ 
and $|{\sf L} \rangle$ describe a photon traveling either in the ``right'' fiber 
or in the ``left'' fiber. How any desired superposition of such two-qubit 
photon states can be prepared is described in Ref.~\cite{Englert}.

In the next section we summarise the basic idea that can be used to construct direct 
confidential quantum communication schemes. In Section 3 we describe concrete 
proposals for its realisation. The security 
against eavesdropping attacks of the general intercept-resend kind is addressed 
in Section 4. After discussing a possible experimental setup, we 
conclude with a summary of our results.
\vspace*{0.5cm}

\noindent{\bf 2. The basic idea}
\vspace*{0.5cm}

In this section we describe the essential ingredients needed for quantum 
cryptography. The basic protocol that all schemes have in common is  
the following: Alice (the sender) exchanges single qubits
with Bob (the receiver), each of them prepared in a certain state. 
As usual she chooses at random between different types of states labeled by $n$. 
Here we choose the notation such that states transmitting a ``$+$'' 
bit are denoted by $|n_+\rangle$. In order to transmit a ``$-$'' bit Alice 
prepares the state $|n_-\rangle$. If a photon arrives at Bob's end, he 
performs a measurement on its state whereat he switches at random between at 
least two different measurement bases. In the following, there will always be only two 
different measurements he can choose and we denote the corresponding basis states 
by $|B_n\rangle$ or $|C_n\rangle$.

As in Refs.~\cite{Letter,book}, let us call the eavesdropper {\em Evan}. 
He has full access to all communication channels between the two parties. This means, 
he can perform any possible quantum mechanical operation on the photons in 
transmission and he can listen into the classical communication between Alice 
and Bob. Security of the scheme is assured when Evan's presence leads to a 
significantly increased error rate in the bit transmission. At the end of a 
transmission, Alice and Bob compare some of their bits to test whether
this rate is above a certain percentage limit or not. If not, Alice announces 
the encrypted message via a classical communication channel or even publicly.

Up to now, our description applies to any quantum cryptography scheme. 
If Alice and Bob want to have a scheme to establish a shared secret key, 
the states $|n_\pm \rangle$, $|B_n \rangle$ and $|C_n\rangle$ only have
to fulfill the condition that Bob can, at least in some cases, deduce which 
bit Alice sent from the knowledge of $n$ and the corresponding outcome of his 
measurement. In BB84, this applies to $50 \, \%$ of the photons which then provide
one key bit each. Whether Evan can gain any knowledge 
about the bits in transmission or not does not matter. Alice and Bob only use 
the key sequence they created if they can verify the absence of any eavesdropping 
attempts.

To obtain a more efficient scheme, Alice and Bob should maximise the rate of 
photons they can use to establish a key bit. One can even assure that Bob 
{\em always} knows whether Alice sent a ``$+$'' or a ``$-$'' bit by using a four 
dimensional Hilbert space \cite{Letter,book}. This is the case, when a photon 
in $|n_+\rangle$ cannot cause the same measurement outcome as a photon in 
$|n_-\rangle$. For instance, if $n=3$ and Bob found $|C_2\rangle$ and knows 
that $|C_2\rangle$ overlaps with the state $|3_+\rangle$, but not with 
$|3_-\rangle$, then he obtains a ``$+$'' bit. Thus every one of Bob's measurements 
matches with whatever state Alice prepared and the scheme is deterministic. 

If Alice and Bob want to communicate directly and confidentially, then 
there is another condition that has to be fulfilled: Whatever operation Evan 
performs on the photon state, he should not be able to gain any 
information about the bit in transmission.  
Let us assume that Alice uses all state pairs with the 
same frequency. To find out whether a photon carries a ``$+$'' or a ``$-$'' bit, 
Evan has to answer the question whether its state belongs to the subspace spanned 
by all $|n_+ \rangle$ states or to the subspace spanned by all $|n_- \rangle$ 
states. If 
\begin{equation} \label{cond}
\sum_n |n_+\rangle \langle n_+| = \sum_n |n_-\rangle \langle n_-| ~,
\end{equation}
then these two subspaces are completely indistinguishable and the bit in 
transmission is perfectly concealed in the state space. 
\vspace*{0.5cm}

\noindent{\bf 3. A concrete scheme for direct communication}
\vspace*{0.5cm}

In this section we present a concrete scheme for direct communication between 
Alice and Bob. To do so let us assume that
\begin{equation} \label{def}
|n_+ \rangle \equiv |B_n \rangle ~~{\rm and}~~ |n_- \rangle \equiv |C_n \rangle ~.
\end{equation}
Then the states $|n_+ \rangle$ and $|n_- \rangle$, respectively, evenly span the 
whole Hilbert space and clearly fulfill condition (\ref{cond}).
They equal either one or the other measurement basis of Bob. 
Note that such a coding is different to BB84 and its modifications, where
where the states $|n_+\rangle$ and $|n_-\rangle$ always belong to the same set of 
basis states.

To assure that Bob always knows how to interpret his measurement result there 
should be no overlap between basis states with the same index $n$, i.e.
\begin{equation} \label{ortho}
\langle B_n | C_n \rangle = 0 ~.
\end{equation}
If Bob finds the photon, for instance, in $|B_m\rangle$ with $m \neq n$, then he 
knows immediately that Alice prepared it $|n_-\rangle$. The reason is that a photon 
in $|n_+\rangle$ cannot cause a ``click'' at this detector. Otherwise, if $n$ 
coincides with $m$, then he knows that he received a ``$+$'' bit. This tells him 
that he measured the same basis as the one used by Alice to prepare the photon state.

The protocol for direct and confidential communication originating from 
this ansatz is the following: First, Alice creates a random succession of ciphers $n$
that will serve as her cryptographic key. The length of this sequence should coincide 
with the length of her message. Depending on whether she wants to transmit a ``$+$'' 
bit or a ``$-$'' bit next, she prepares the photon either in $|B_n \rangle$ or in 
$|C_n \rangle$ with $n$ according to the next number of her key and sends 
it to Bob. Bob measures at random either the $B$ or the $C$ basis on each incoming 
photon. After Alice and Bob assured each other that the transmission was 
secure (how well this can be done is discussed in the next section), Alice publicly 
announces her key. In doing so, she reveals the message to Bob.

Up to now, we have not yet answered the question, what the $B$ and $C$ basis 
should look like. Let us assume here that Alice and Bob use single photon 
two-qubit states \cite{Englert}. Then all states are part of a four-dimensional Hilbert 
space but our results can also be carried over easily to higher dimensions. Note,
that is not possible to find a non-trivial solution to Eqs.~(\ref{def}) and (\ref{ortho}) 
in less than four dimensions, this means a solution for which 
the states $|C_n\rangle$ are not just a permutation of the states of the $B$ basis.

In the following, we denote the basis transformation that rotates the $B$ basis into 
the $C$ basis by ${\bf A}$ and write
\begin{equation} \label{CBA}
\bigl(|C_1\rangle, |C_2\rangle, |C_3\rangle, |C_4\rangle \bigr) 
= \bigl(|B_1\rangle, |B_2\rangle, |B_3\rangle, |B_4\rangle \bigr) \, {\bf A} ~. 
\end{equation}
Condition (\ref{ortho}) is then fulfilled if the unitary $4\times 4$ matrix ${\bf A}$ 
has only vanishing diagonal elements. Besides this, there are no other restrictions 
on ${\bf A}$ and there are many choices Alice and Bob can make. For symmetry reasons, 
let us assume that ${\bf A}$ is not only unitary but also Hermitian. Then the inverse of 
the transformation (\ref{CBA}) is also furnished by ${\bf A}$. 
 
It is sufficiently general to consider matrices of the form
\begin{equation} \label{Agen}
{\bf A} = {\rm i} \left( \begin{array}{rrrr}
 0   &  a_1 &  a_2 &  a_3 \\
-a_1 &  0   &  a_3 & -a_2 \\
-a_2 & -a_3 &  0   &  a_1 \\
-a_3 &  a_2 & -a_1 &  0 \end{array} \right) 
\end{equation}
where the parameters $a_i$ are real and fulfill the normalisation constraint
\begin{equation} 
a_1^2+a_2^2+a_3^2 =1~.
\end{equation}
Thus, Alice and Bob have two free parameters which they can choose to their liking.
Bob's probabilities to find the incoming photon in a certain state are summarised 
in Table \ref{tbl}. 

\begin{table}
\begin{center}
\caption{\label{tbl}%
Bob's probability to find a certain measurement outcome for the scheme corresponding 
to Eq.~(\ref{Agen}). \vspace*{0.5cm}}
\begin{tabular}{c|ccccccccc}
\hline \hline \rule{0pt}{12pt} 
Alice's & \multicolumn{9}{c}{measurement outcome at Bob's end} \\
photon state & $B_1$ & $B_2$ & $B_3$ & $B_4$ && $C_1$ & $C_2$ & $C_3$ & $C_4$~
\\ \hline \rule{0pt}{12pt} 
$1_+$ & $1$ & $0$ & $0$ & $0$ && $0$ & $a_1^2$ & $a_2^2$ & $a_3^2$ \\
$2_+$ & $0$ & $1$ & $0$ & $0$ && $a_1^2$ & $0$ & $a_3^2$ & $a_2^2$ \\
$3_+$ & $0$ & $0$ & $1$ & $0$ && $a_2^2$ & $a_3^2$ & $0$ & $a_1^2$ \\
$4_+$ & $0$ & $0$ & $0$ & $1$ && $a_3^2$ & $a_2^2$ & $a_1^2$ & $0$ 
\\ \hline \rule{0pt}{12pt} 
$1_-$ & $0$ & $a_1^2$ & $a_2^2$ & $a_3^2$ && $1$ & $0$ & $0$ & $0$ \\
$2_-$ & $a_1^2$ & $0$ & $a_3^2$ & $a_2^2$ && $0$ & $1$ & $0$ & $0$ \\
$3_-$ & $a_2^2$ & $a_3^2$ & $0$ & $a_1^2$ && $0$ & $0$ & $1$ & $0$ \\
$4_-$ & $a_3^2$ & $a_2^2$ & $a_1^2$ & $0$ && $0$ & $0$ & $0$ & $1$ 
\\ \hline \hline
\end{tabular}
\end{center}
\end{table}

In the next two Sections we will see that the fully symmetric choice
\begin{equation} \label{Aopt}
a_1 = a_2 = a_3 = 1/\sqrt{3} 
\end{equation}
maximises the error rate that a potential eavesdropper 
introduces in the bit transmission between Alice and Bob. Another 
solution is to choose
\begin{equation} \label{Asimp}
a_1 = a_2 = 1/\sqrt{2} ~~ {\rm and} ~~ a_3=0 ~. 
\end{equation}
In this case, the experimental implementation of the corresponding scheme is 
particularly simple, see in Section 5 below. 
To prepare a photon in one of the states of the $B$ and the $C$ basis 
does not require to produce entanglement between the two degrees of freedom, the 
spatial coordinates and the polarisation, of the photon. Nevertheless, the error 
rate introduced by an eavesdropper is still relatively large.

Finally, we would like to convince ourselves that whatever the parameters $a_i$ are, 
it is indeed impossible for Evan to gain any information without the knowledge 
of Alice's key. The reason is that the states $|n_+ \rangle$ (and $|n_- \rangle$,
respectively) equally span the whole Hilbert space. As long as Alice chooses 
equally likely between the four possible values of $n$, she prepares
the ensemble of photons with a ``$+$'' bit in the mixed state that is given 
by the (normalised) identity matrix. The same applies to the ensemble of photons 
that carry a ``$-$'' bit. Thus whatever Evan measures, 
the probability to find the photon in a certain state always equals ${1 \over 4}$ 
and finding a certain state does not reveal any information to Evan.
\vspace*{0.5cm}

\noindent{\bf 4. Security against intercept-resend attacks}
\vspace*{0.5cm}

The security of the scheme we present here results from the fact that 
Alice does not reveal her key before she is not convinced that no 
eavesdropper has been listening in. To test whether this 
is the case or not, Alice and Bob proceed as follows: Alice intersperses her 
message with a fair number of control bits at random positions and of random 
values. Only Alice knows which ones are the control bits and which ones 
the message bits. After the transmission of all photons, she tells Bob which 
photons carried control bits and he tells her in which state he found them. 
If Alice verifies that Bob's findings are consistent with what she sent,
then they conclude that the transmission was secure. Otherwise, if the error 
rate is above a certain percentage level, they should not trust in the security 
of their communication and Alice should repeat her 
transmission using a different cryptographic key. 

Let us now imagine that Evan is listening in and determine the minimum error 
rate he causes in case of the setup described in the previous section. 
By doing so we do not care whether Evan can gain any information in this way 
or not, once Alice reveals her key. Let us assume, as usual, that Evan 
intercepts every photon and performs a measurement on it. Afterwards he 
forwards a replacement to Bob accordingly, namely in the two-qubit state that has 
the best chance of avoiding wrong detector clicks at Bobs end. 

In the following we denote the states of Evan's measurement basis by $|E_k \rangle$.
As explained at the end of the previous section, the probability to find the 
photon in a certain state always equals ${1 \over 4}$ and Evan cannot gain any 
information from his measurement. For simplicity we consider only the 
strategy in which Evan forwards the photon in exactly the same state he found 
it in. In this way he forwards it at least in a state that has 
some overlap with the state prepared by Alice. More general strategies in which 
Evan optimises the forwarded state can be analysed as well, but that is technically 
more demanding, and we are here content with referring the reader to the detailed 
discussion in \cite{book}.

The probability that Alice prepares her photon in $|n_+\rangle$ equals 
${1 \over 8}$ and the probability that Bob measures the $B$ basis is given by 
${1 \over 2}$. In this case, an error occurs if Bob finds the photon in 
$|B_m\rangle$ with $n \neq m$. The contribution of this case to the total
error probability is given by 
\begin{equation}
{\textstyle {1 \over 8}} \cdot {\textstyle {1 \over 2}} \, \sum_k
|\langle E_k|n_+\rangle|^2 \sum_{m \neq n} |\langle E_k|B_m \rangle|^2 
= {\textstyle {1 \over 16}} \, 
\Big[ \, 1- \sum_k |\langle E_k|B_n \rangle|^4 \, \Big]  
\end{equation}
because the probability that Evan measures $|E_k\rangle$ is in this case 
$|\langle E_k|n_+\rangle|^2$ whilst $|\langle E_k|B_m \rangle|^2$ is the 
probability that Bob finds the photon afterwards in $|B_m \rangle$. 
Analogously, one finds that the contribution to the error rate is given by 
\begin{equation} \label{10}
{\textstyle {1 \over 16}} \sum_k |\langle E_k|n_+\rangle|^2 \, 
|\langle E_k|n_- \rangle|^2 
= {\textstyle {1 \over 16}} \sum_k |\langle E_k|B_n\rangle|^2 
\, |\langle E_k|C_n \rangle|^2 ~, 
\end{equation}
if Bob measures the $C$ basis instead of the $B$ basis. In this case an error 
occurs only if Bob finds the photon in $|n_-\rangle$. Calculating the 
contributions to the total error rate when Alice prepares the photon in the 
state $|n_-\rangle$ leads to the same result but with $|B_n \rangle$ replaced 
by $|C_n \rangle$ and vice versa. 

To calculate the total error rate $P_{\rm error}$ one has to sum over all 
contributions and all possible values of $n$. Doing so leads to
\begin{eqnarray} \label{Pk}
P_{\rm error}
&=& \sum_n \Big[ \, {\textstyle {1 \over 8}}
- {\textstyle {1 \over 16}} \sum_k \Big( \, |\langle E_k|B_n \rangle|^4  
+ |\langle E_k|C_n \rangle|^4 + 2 \, |\langle E_k|B_n \rangle|^2 
\, |\langle E_k|C_n \rangle|^2 \, \Big) \Big] \nonumber \\
&=& {\textstyle {1 \over 2}} -  {\textstyle {1 \over 16}} \sum_n \sum_k 
\Big( \, |\langle E_k|B_n \rangle|^2 - |\langle E_k|C_n \rangle|^2 \, \Big)^2 ~.
\end{eqnarray}
Evan's task of minimising the error rate so reduces to the task of minimising 
this expression. Using the notation
\begin{equation} 
|E_k \rangle = \sum_m e_m \, |m_+ \rangle  
\end{equation}
and Eqs.~(\ref{def})--(\ref{Agen}) we find that
\begin{equation} 
P_{\rm error} \ge 
{\textstyle {1 \over 2}} - {\textstyle {1 \over 16}} 
\sum_n \big( \, 1+ a_1^4 + a_2^4 + a_3^4 \, \big) \sum_m |e_m|^4 
\end{equation}
by neglecting all negative terms in the round brackets at the right hand side of 
Eq.~(\ref{Pk}), that is: the terms that stem from Eq.~(\ref{10}). 
The state $|E_k \rangle$ is normalised and its coefficients $e_m$
obey the inequality $\sum_m |e_m|^4 \le \sum_m |e_m|^2=1$. This leads to the result
\begin{equation} \label{error}
P_{\rm error} \ge  
{\textstyle {1 \over 4}} \, \big( \, 1- a_1^4 - a_2^4 - a_3^4 \, \big) ~.
\end{equation}
For the optimal scheme corresponding to the parameters given in Eq.~(\ref{Aopt})
the right hand side of this equation is ${1 \over 6} = 16.67 \, \%$. For the 
scheme (\ref{Asimp}), the error rate introduced by Evan in the bit transmission 
is always above ${1 \over 8} = 12.5 \, \%$.

\begin{center}
\begin{figure}
\epsfig{file=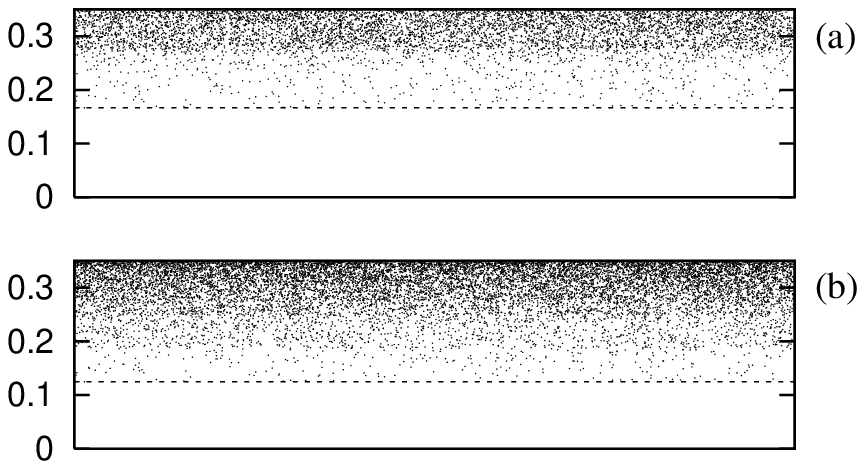,width=14.5cm} \\[0.2cm]
\caption{Error rate introduced by Evan into the bit transmission between Alice 
and Bob for the parameters chosen (a) as in Eq.~(\ref{Aopt}), where it is 
always above $16.67 \, \%$, and (b) for the parameters as in Eq.~(\ref{Asimp})
where it is always above $12.5 \,\%$. Each point corresponds to a different 
intercept-resend strategy whereat Evan's measurement basis and the state in 
which he forwards the photon to Bob have been acquired completely 
at random.}\label{fig1}
\end{figure}
\end{center}

As it stands, this calculation applies only to strategies where Evan forwards the 
photon in the detected state, but not to those where the forwarded states are 
optimised for minimal error rates. It turns out, however, that these more 
sophisticated strategies do not yield error rates below these $16.67 \, \%$ or 
$12.5 \, \%$, respectively. This is confirmed by the numerical data presented in 
Figure \ref{fig1} which reports error rates for random choices of Evan's 
measurement basis and forwarded states. The data demonstrate 
that the right hand side of Eq.~(\ref{error}) is indeed the lower bound 
of the error rate. 

To derive Eq.~(\ref{error}) we neglected only terms 
proportional $|e_k|^2 |e_j|^2$ with $k \neq j$. An optimal strategy for Evan 
is therefore, for instance, to measure the $B$ basis, i.e.~to measure whether 
the incoming photon is in one of the states $|n_+ \rangle$. 
This strategy also optimises eavesdropping with respect to maximising 
the information gain of Evan as soon as he gets to know the key. If
Alice and Bob fail to notice his presence, he can intercept the whole 
transmitted message.
\vspace*{0.5cm}

\noindent{\bf 5. Proposal for an experimental realisation}
\vspace*{0.5cm}

We have seen in the previous section that the error rate introduced by an 
eavesdropper is always above $16.67 \, \%$ for the optimal choice of the 
parameters (\ref{Aopt}). This is not much larger than the minimum error rate of 
$12.5 \, \%$ which was found for the parameter choice of Eq.~(\ref{Asimp}). 
In this section we discuss how the second scheme could be realised experimentally
because its implementation is much simpler, although implementing the optimal 
scheme is also possible with the methods of \cite{Englert}. 
To achieve the same degree of security in the second scheme, Alice and Bob must use 
about $40\,\%$ more control bits.

Let us assume now that the vectors of the $B$ basis are given by
\begin{equation} \label{B}
\bigl(|B_1\rangle, |B_2\rangle, |B_3\rangle, |B_4\rangle \bigr) 
= \bigl({\sf |Rv \rangle, |Lv \rangle, |Lh \rangle, |Rh} \rangle \bigr) ~.
\end{equation}
Then the states of the $C$ basis equal
\begin{equation} \label{C}
\bigl(|C_1\rangle, |C_2\rangle, |C_3\rangle, |C_4\rangle \bigr) 
= {\rm i} \, \bigl({\sf -|Ls \rangle, |Rs \rangle, |Ra \rangle, -|La} \rangle \bigr) ~,
\end{equation}
where 
\begin{equation} \label{sym-asym}
|{\sf s}\rangle \equiv \textstyle{{1 \over \sqrt{2}}} \,  
\bigl(|{\sf v}\rangle + |{\sf h}\rangle\bigr) ~~{\rm and} ~~
|{\sf a}\rangle \equiv \textstyle{{1 \over \sqrt{2}}} \, 
\bigl(|{\sf v}\rangle - |{\sf h}\rangle\bigr) 
\end{equation}
are the symmetric and the antisymmetric superposition of the basic polarisation states. 
Thus, the $C$ basis differs from the states of the $B$ basis only with respect to the 
possible polarisations of the photons. The phase factors $\pm {\rm i}$ in Eq.~(\ref{C})
are a consequence of the conventions adopted at (\ref{Agen}) and (\ref{Asimp}) 
and could as well be omitted. To prepare the states $|n\pm\rangle$, Alice 
could use any source that produces single photons on demand. For examples of 
experimental realisations of such sources see for instance 
Refs.~\cite{Yamamoto,Kurtsiefer,Kuhn,Lounis}.

If Alice wants to send a ``$+$'' bit to Bob, then she should prepare the photon at 
random in one of the four states on the right hand side of Eq.~(\ref{B}). To do so 
she chooses equally likely between the polarisations $|{\sf h} \rangle$ and 
$|{\sf v} \rangle$ and sends the photon either through a ``left'' or a ``right'' 
fiber. To send a ``$-$'' bit, Alice can proceed in the same way but 
should then change the polarisation of the outgoing photon before sending it to 
Bob. This can be done, for instance with the help of a half-way plate (HWP)
that affects the polarisation of a photon such that $|{\sf s}\rangle$ changes 
into $|{\sf v}\rangle$ and $|{\sf a}\rangle$ changes into $|{\sf h}\rangle$, or 
vice versa. For practical realisations, electrically controllable Pockels-cells
should be used. 

\begin{center}
\begin{figure}
\epsfig{file=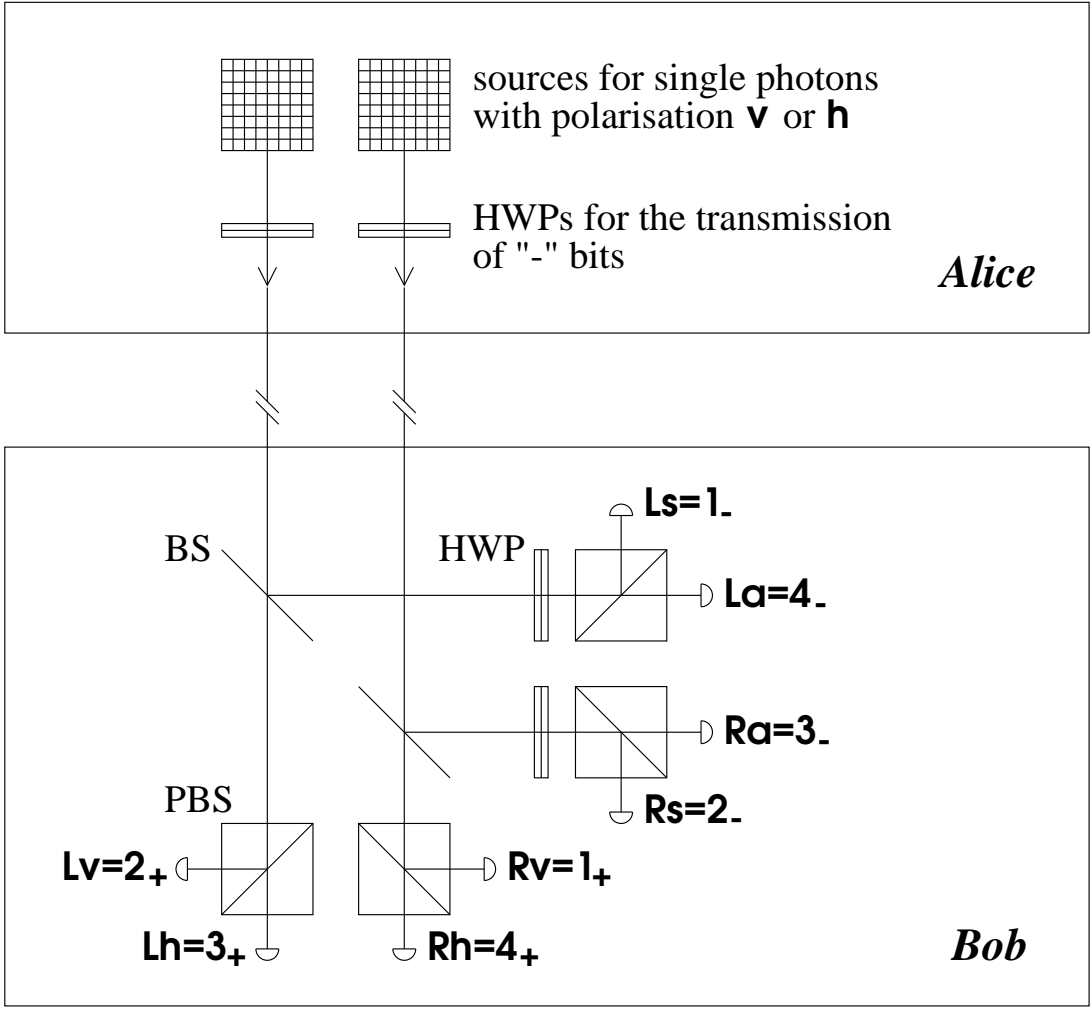,width=14.5cm} \\[0.2cm]
\caption{A feasible experimental setup for the transmission of a ``$-$'' bit with 
single-photon two-qubit states. For an explanation of the optical elements see 
text. To transmit ``$+$'' bits, the HWPs have to be turned around so that they do not
change the polarisation of the outgoing photon. Otherwise, the experimental setup is 
exactly the same.}\label{fig3}
\end{figure}
\end{center}

A possible experimental setup for the transmission of ``$-$'' bits is 
sketched in Figure \ref{fig3}. To deflect vertically polarised photons to one detector 
and horizontally polarised photons to another detector, Bob uses polarising 
beam splitters (PBS) whilst he changes the polarisation of a photon, like
Alice, with the help of a HWP.  
In which state Bob finds a photon in case of a ``click'' is indicated in 
Figure \ref{fig3} by the two letters written next to the corresponding detector. 
At Bob's end, a beam splitter (BS) reroutes the photon either to a measurement of 
the $B$ or the $C$ basis.

Instead of using a ``left'' and a ``right'' fiber and two polarisation degrees 
of freedom, Alice and Bob could also utilise other parameters to create single photon 
two-qubit states. The two fibers can, for instance, be replaced by one fiber 
and Alice and Bob agree for each photon about two small time windows around a 
time $t_{\sf L}$ and a time $t_{\sf R}$. If Alice sends the photon around 
$t_{\sf L}$ it means that she prepared it in the state $|{\sf L} \rangle$, otherwise,
if she sends the photon around $t_{\sf R}$, she prepared it in $|{\sf R} \rangle$.
Alternatively, two degrees of freedom could also be obtained by exploiting different 
photon frequencies. 

The scheme shown in Figure \ref{fig3} looks as if it were a combination of two BB84 
schemes. But in fact it is not. The scheme is more efficient than what one would get by just 
combining two BB84 schemes naively. The reason is that the results are 
interpreted in a completely different way (see Section 3 and Table \ref{tbl}).
In contrast to BB84, each and every photon sent by Alice transmits one bit 
and the scheme is therefore deterministic. In addition, the bit 
in transmission is concealed in such a way that an eavesdropper cannot gain any information 
by performing measurements on the photon state (which cannot be achieved in BB84), 
and Alice and Bob can communicate directly and confidentially.
\vspace*{0.5cm}

\noindent{\bf 6. Conclusions}
\vspace*{0.5cm}

In summary, we discussed a new scheme for direct and confidential communication 
between Alice and Bob in detail. While both parties exchange single bits 
(carried by photons) as in any other quantum cryptography scheme, the purpose 
of the bit transmission is completely different. Instead of establishing a shared 
secret key, which can be used later to encrypt a message, Alice can send her message 
directly. What requirements such schemes have to meet in general has been
discussed in Section 1. 

To encrypt her message, Alice creates a random sequence of ciphers -- the
cryptographic key. To transmit a ``$+$'' bit she prepares the photon in the state
$|n_+\rangle$, to transmit a ``$-$'' bit she prepares it in $|n_-\rangle$,
whereat $n$ always coincides with the next cipher of her key. After Alice and Bob 
verified that no eavesdropper was listening in, Alice can publicise her key
without hesitation. She knows that Bob will be the only one who can decode her 
message, because he is the only one who received it. In this sense, the scheme 
realises quantum cryptography with a {\em publicly known} key. 

In Section 2 and 3 we presented a concrete protocol based on single-photon two-qubit 
states and discussed its security against intercept-resent eavesdropping strategies. 
As in other schemes, security arises from the fact that the presence of an 
eavesdropper leads to a significantly increased error rate in the bit transmission.
This rate can be determined by comparing some control bits with which Alice 
had interspersed the message before. Only when the measurement outcomes of Bob's 
side match with the states in which Alice prepared the control qubits, 
both parties should trust in the security of their communication and Alice can 
announce her cryptographic key.

By choosing the parameters that characterise the scheme suitably, Alice and
Bob can assure that the error rate Evan introduces in the bit transmission
is always above $16.67 \, \%$. Nevertheless, in Section 5 we discussed  
possibilities for the experimental realisation of another scheme, 
one in which the error rate can be as low as $12.5 \, \%$. 
The advantage of this scheme is that its implementation is much simpler, 
although implementing the optimal scheme is possible too.

\vspace*{0.5cm}
{\em Acknowledgment}.
B.-G. E. thanks Yakir Aharonov and Lev Vaidman for illuminating discussions and 
wishes to express his gratitude for the hospitable environment provided by Gerald 
Badurek and Helmut Rauch at the Atominstitut in Vienna, where part of this work 
was done. A. B. and B.-G. E. are grateful for the kind hospitality extended to them 
at the Erwin-Schr\"odinger-Institut in Vienna. Ch. K. and H. W. acknowledge support 
by project QuCommm (IST-1999-10033) of the European Union.

\newpage
\noindent

\noindent

\newpage
\begin{center}
  ERRATUM
\end{center}

\noindent%
\textbf{A. Beige, B.-G. Englert, Ch.\ Kurtsiefer and H. Weinfurter},
Secure Communication with a Publicly Known Key,
\textit{Acta Phys.\ Pol.\ A\/} \textbf{101}, 357 (2002).

\bigskip

\noindent%
The particular communication scheme that is specified by the parameter choice 
of Eqs.~(8) and described in Fig.~2 is not secure. 
It falls prey to an eavesdropping attack in which 
a quantum nondemolition (QND) measurement
is performed that distinguishes $\mathsf{L}$ from $\mathsf{R}$.
An intermediate measurement of this kind does not produce error's 
at Bob's end and would, therefore, not be noticed. 
But it gives Evan just enough information to infer correctly
the value of every bit after Alice publishes her key sequence.

The main conclusions of the paper are not affected by this flaw.
The generic communication scheme, in which all three $a_i$ parameters 
of Eq.~(5) are nonzero, \emph{is} secure, because 
there are no nontrivial QND measurements unless one of the $a_i$'s vanishes.
Any attempt by Evan to acquire the value of every bit sent will
unavoidably give rise to errors, with the total error rate correctly 
stated in Eq.~(14).

We are greatly indebted to Daniel Collins for detecting the flaw and telling
us about it.

\end{document}